\documentclass[aip,jcp,reprint,showkeys,superscriptaddress,a4paper,amsmath]{revtex4-1}
\usepackage[pdftex]{graphicx}
\usepackage{epsfig}
\usepackage{amsmath}
\usepackage{graphics}
\usepackage{latexsym}
\usepackage{amssymb}
\usepackage{inputenc}
\usepackage{color}
\usepackage{siunitx}
\usepackage[colorlinks=true,linkcolor=magenta,citecolor=cyan]{hyperref}

\newcommand{\la}{\left\langle}
\newcommand{\ra}{\right\rangle}

\newcommand{\angstrom}{\textup{\AA}}

\begin{document}

\title{Correlation between ordering and shear thinning in confined liquids}

\author{Yusei Kobayashi}
\affiliation{Department of Mechanical Engineering, Keio University, Yokohama, Kanagawa 223-8522, Japan}
\email{kobayashi@mech.keio.ac.jp}
\author{Noriyoshi Arai}
\affiliation{Department of Mechanical Engineering, Keio University, Yokohama, Kanagawa 223-8522, Japan}
\email{arai@mech.keio.ac.jp}
\author{Kenji Yasuoka}
\affiliation{Department of Mechanical Engineering, Keio University, Yokohama, Kanagawa 223-8522, Japan}
\email{yasuoka@mech.keio.ac.jp}

\begin{abstract}
Despite the extensive research that has been conducted for decades on the behavior of confined liquids, detailed knowledge of this phenomenon, particularly in the mixed/boundary lubrication regime, remains limited. This can be attributed to several factors including the difficulty of direct experimental observations of the behavior of lubricant molecules under non-equilibrium conditions, the high computational cost of molecular simulations to reach steady state, and the low signal-to-noise ratio at extremely low shear rates corresponding to actual operating conditions. To this end, we studied the correlation between the structure formation and shear viscosity of octamethylcyclotetrasiloxane confined between two mica surfaces in a mixed/boundary lubrication regime. Three different surface separations corresponding to two-, three-, and five-layered structures were considered to analyze the effect of confinement. The orientational distributions with one specific peak for $n=2$ and two distributions, including a parallel orientation with the surface normal for $n>2$, were observed at rest. The confined liquids exhibited a distinct shear-thinning behavior independent of surface separations for a relatively low sliding velocity, $V_{\rm x}\lesssim 10^{-1}\,{\rm m/s}$. However, the shear viscosities at $V_{\rm x}\lesssim 10^{-1}\,{\rm m/s}$ depended on the number of layered structures. Newtonian behavior was observed with a further increase in the sliding velocity. Furthermore, we found a strong correlation between the degree of molecular orientation and the shear viscosity of the confined liquids. The magnitude of the shear viscosity of the confined liquids can primarily be determined by the degree of molecular orientation, and shear-thinning originates from the vanishing of specific orientational distributions with increasing sliding velocity.
\end{abstract}
\maketitle

\section{Introduction}
Energy-saving techniques such as friction reduction have become increasingly important as countermeasures against global warming.\cite{Luo:ne:2021}
Particularly, an improvement in the friction reduction\cite{Hod:nat:2018} for mechanical systems, including automobile systems, is essential because the internal combustion engine is still used as the primary power source for most vehicles, and various tribological contacts such as electric motors, wheel bearings, and steering systems are still present in electric vehicles.\cite{Farfan-Cabrera:ti:2019}
In the analysis of lubrication problems, the behavior of the lubricant molecules in a wide range of fluid lubrication to boundary lubrication must be considered based on the practical operating conditions of mechanical systems (e.g., engine oil experiences an engine starting/idling state (boundary lubrication) and a high-speed operating state (fluid lubrication)).
The Stribeck curve\cite{Stribeck:zvsi:1902} is a fundamental concept of tribology.
This concept describes the friction coefficient as a function of the Sommerfeld number (viscosity $\times$ sliding velocity/load).
In the fluid lubrication region, the frictional force is determined by the bulk properties of the lubricant.
Therefore, the use of a lower bulk viscosity can effectively reduce the friction force.
A transition from fluid lubrication to a mixed/boundary lubrication region is observed with the decrease in the sliding velocity and/or increase in the load, and lubricant films are formed under extreme confinement at the nanometer scale.
Therefore, obtaining a better understanding of the bulk properties along with the structural and shear behavior of the lubricant under nanoscale confinement is crucial.

Confined liquids have attracted considerable attention in the past few decades owing to their potential applications in electronics\cite{Gao:nc:2019}, engineering\cite{Cafolla:sa:2020}, and biomedical\cite{Lin:amr:2022} fields.
The molecular arrangement and orientation in the static state, and the dynamic properties can be measured on a length scale from $\si{\micro}$m to nm owing to the progress in experimental techniques such as atomic force microscopy (AFM)\cite{Cafolla:ns:2018}, surface force apparatus (SFA)\cite{Kimura:pj:2021}, and resonance shear measurement (RFM)\cite{Mizukami:la:2022}.
Confined liquids exhibit distinct phases\cite{Gao:sr:2018,Jiang:jacs:2021} and flow properties\cite{Nair:sci:2012,Sun:cpl:2021} not exhibited in the bulk solution even for low-molecular-weight solutions such as water owing to the effect of spatial constraints and solid-liquid interfaces.
Confined water is an interesting system because it is ubiquitous in nature; examples include biochannels\cite{Marbach:np:2018,Arai:ns:2020} and water films confined between clay mineral surfaces\cite{Yoshida:nc:2018,Chang:jpse:2021}.
Pure water and lubricating oil with trace quantities of water are also potential candidates for future lubricants due to their eco-friendly nature.\cite{Liang:ass:2019,Cafolla:ns:2020,Sun:cpl:2021}
A recent experiment\cite{Cafolla:ns:2020} conducted on hexadecane with trace quantities of water confined between mica walls demonstrated that the nucleation of water nanodroplets at the surface is induced by an increase in the temperature of the system, and the addition of amphiphilic molecules such as oleic acids to the lubricant suppresses this nucleation.
Additionally, research conducted on the viscosity behavior of four different phenyl ether lubricants under nanoslit confinement using RSM indicated a significant increase in the viscosity with the decrease in the surface separation ($\lesssim 2\,{\rm nm}$), and the magnitude of the viscosity correlation between the bulk and the confined systems was reversed.\cite{Watanabe:tl:2014}
Despite the extensive research conducted on the development of measurement techniques for the properties of confined lubricants, obtaining a better understanding of the correlation between the structure and flow properties of lubricants under nanoconfinement remains a challenge owing to the difficulty of obtaining direct experimental observations of the behavior of lubricant molecules, particularly under non-equilibrium conditions.

Molecular dynamics (MD) simulation is a powerful tool which can be used to obtain microscopic insights into the properties of confined liquids behind the experimental results and to guide future research applications.
Several studies have been conducted using both equilibrium and non-equilibrium MD simulations on confined liquids such as Lennard-Jones (LJ) particles\cite{Kaneko:cpl:2010,Doi:aipa:2017,Sheibani:fpe:2020}, water\cite{Sharma:pnas:2012,Gao:sr:2018,Zhang:rsca:2019,Sun:cpl:2021}, and hydrocarbons\cite{Ta:ti:2017,Krass:tl:2018,Gao:lub:2022}.
A recent simulation conducted by Ta $et$ $al.$\cite{Ta:ti:2017} of a hexadecane film confined between iron oxide surfaces demonstrated that an increase in the shear rate can result in a decrease in the degree of ordering of the hexadecane film and an increase in the velocity slip at the solid-liquid interfaces.
A combination of tribometer experiments and MD simulations\cite{Ewen:pccp:2017} demonstrated the effect of molecular structure on the friction behavior.
The study reported that flexible and broadly linear hydrocarbon molecules exhibit low friction coefficients which increase with the strain rate and pressure, whereas inflexible molecules that contain cycloaliphatic groups have high friction coefficients that are almost independent of the strain rate and surface pressure.
However, the task of computing non-equilibrium MD simulations in the shear rate range corresponding to those used in practical operating conditions is extremely difficult, which hinders the possibility of obtaining a detailed understanding of the underlying physics despite the extensive simulation research conducted on this topic.
Therefore, most of the previous simulation efforts have been focused on friction and flow behavior in the elastohydrodynamic lubrication regime.

This study aims to provide new molecular-level insights into the correlation between the structural and lubrication properties in the mixed/boundary lubrication regime.
Octamethylcyclotetrasiloxane (OMCTS) and cyclohexane are among the most widely used lubricant molecules in practical lubrication systems; these model lubricants are more frequently used in tribometer experiments because of their globular and non-polar nature.
Therefore, these molecules have also been used as model lubricants in molecular simulation studies conducted on confinement.
Hitherto, force-field parameters for OMCTS\cite{Matsubara:jctc:2010,Matsubara:jcp:2011,Xu:jcp:2014} and cyclohexane\cite{Jorgensen:jacs:1996,Cummings:aj:2010} have been developed to reproduce the layered structure and period of oscillatory force observed in experiments.
In addition to the development of potential models, analyses of the transport properties\cite{Matsubara:prl:2012} and the possibility of freezing transition\cite{Vadhana:jpcb:2016} under nanoscale confinement have been conducted $via$ equilibrium MD simulations.
A confinement-induced diffusion slowdown was observed for OMCTS confined between mica surfaces, and the mechanism of this phenomenon was explained using the concept of activated diffusion theory\cite{Matsubara:prl:2012}.
As an exception, a recent simulation of cyclohexane confined between mica surfaces by Xu $et$ $al.$\cite{Xu:pnas:2018} presented the first molecular evidence for the structural changes and dynamics of stick-slip motion in boundary lubrication.

In this study, we performed MD simulations to demonstrate the correlation between structure formation and shear viscosity of a lubricant film under nanoscale confinement in a mixed/boundary lubrication regime.
We focused on the OMCTS-mica system, given the progress made in lubricant models to accurately reproduce a realistic view of the microscopic structure of the confined system.
By using this system, our work determines the physical mechanism of the behavior of confined liquids in the mixed/boundary lubrication regime from a molecular viewpoint.

\section{Simulation methods}
\label{sec:method_MD}
\subsection{Simulation model}
We use the MD method to analyze the structural formation and flow properties of OMCTS confined between two mica walls in equilibrium and under shear.
In the employed OMCTS model\cite{Matsubara:jctc:2010}, both the crystal lattice constants and liquid transport properties such as bulk density and diffusion coefficient in the range of 300-440\,K are reproduced, despite only considering the methyl-methyl interactions.
The atomic sites interact with each other through the standard LJ potential, which is given as:
\begin{equation}
	U(r_{ij}) = \begin{cases}
	4\varepsilon_{ij} \left[\left(\frac{\sigma_{ij}}{r_{ij}}\right)^{12}-\left(\frac{\sigma_{ij}}{r_{ij}}\right)^{6}\right],
	\; r_{ij} \leq r_{\rm cut} \\
	0, \;\;\;\;\;\;\;\;\;\;\;\;\;\;\;\;\;\;\;\;\;\;\;\;\;\;\;\;\;\;\;\;\;\;\;\;\; r_{ij} > r_{\rm cut},
	\end{cases}
	\label{eq:ULJ}
\end{equation}
where $r_{ij}$ denotes the distance between atomic sites, $\sigma_{ij}$ is the length parameter, $\varepsilon_{ij}$ is the energy parameter, and $r_{\rm cut} = 10.0\,{\rm \angstrom}$\cite{Matsubara:jcp:2011} is the cutoff radius.
All the LJ parameters used in this study are listed in Table~\ref{tab:Interaction}.
\begin{table}[b]
\caption{Lennard-Jones potential parameters\cite{Matsubara:jctc:2010,Matsubara:jcp:2011}}
\begin{center}
  \begin{tabular*}{0.48\textwidth}{@{\extracolsep{\fill}}lll}
    \hline
    & $\sigma\,({\rm \angstrom})$ & $\varepsilon$\,(kcal/mol)\\
    \hline
    Me--Me           & 3.540 & 0.390\\
    Me--O            & 3.393 & 0.269\\
    Me--${\rm K}^+$  & 2.817 & 2.611\\
    \hline
  \end{tabular*}
  \end{center}
    \label{tab:Interaction}
\end{table}

\subsection{Simulation procedure}
We performed MD simulations using a liquid-vapor (L-V) molecular ensemble proposed by Leng\cite{Leng:jpcm:2008} to prepare the confined system.
The system setup for the L-V MD simulations of the OMCTS-mica system was established based on previous studies\cite{Matsubara:jcp:2011}, as shown in Fig.~S1 in the ESI$^\dag$.
The mica surface consists of 614 O and 288 $\rm {K^+}$ atoms.
The cleaved mica surface contains half of the $\rm {K^+}$ ions present in the crystal.
The 1100 OMCTS molecules were sandwiched between two mica walls with a surface area of $81.8\,{\rm \angstrom} \times 81.2\,{\rm \angstrom}$, and the two mica walls were connected to smooth repulsive walls expressed by harmonic potentials.
The simulation box was $400\,{\rm \angstrom} \times 81.2\,{\rm \angstrom} \times 110\,{\rm \angstrom}$, and the periodic boundary condition was applied in all three spatial conditions.
The complete details of the procedure have been provided in Ref~\cite{Matsubara:jcp:2011}.
This atomistic model for the OMCTS-mica system can reproduce the layered structure and period of oscillatory force observed in the experiments.

We calculated the normal surface pressure (force/area)-distance curve at $1\,{\rm \angstrom}$ intervals by continuously moving the upper mica with $dz/dt=0.01\,{\rm \angstrom}/{\rm ps}$ to obtain confined systems with a targeted number of layers ($n$).
The initial configurations were prepared by randomly inserting OMCTS molecules between two mica walls with $H=63\,{\rm \angstrom}$ using the Packmol software\cite{Martinez:jcc:2009}.
The temperature was maintained at 300 K using a Nos\'{e}-Hoover thermostat\cite{nose:ptps:1991} with a time constant of 0.2\,ps.
The atoms of the upper and lower mica were frozen, and each OMCTS molecule was considered as a rigid body\cite{Kamberaj:jcp:2005} with a time step of 2.0\,fs.
For each surface separation, $H$, an equilibrium run was carried out for at least 1\,ns before performing a production run to calculate the normal force $F_{\rm z}$.
We selected $H$ from the oscillation peak of the target layer region after deriving the normal surface pressure-distance curves from the L-V MD simulations; however, the value of $H$ depends on the degree of surface pressure (see Fig.~S\ref{SMfig:Fcurve} in the Supplementary Materials).
For example, the surface separation for a five-layered system is determined according to the following procedure.
First, we selected $H$ from the oscillation peak of the five-layer region that corresponds to the normal surface pressure, which can be obtained from the surface force experimental measurements.
A five-layered system was obtained at a normal surface pressure, $P_{\rm zz}$, of 0.1\,MPa in the experiments\cite{Mizukami:private:2022}.
Here, we identified two values of $H$ at $P_{\rm zz}=0.1\,{\rm MPa}$ within the oscillation peak of the five-layer region.
Previous MD simulations of the LJ particles confined within a nanoslit have demonstrated that the layered structures at larger values of $H$ within the oscillation peak are energetically more stable, resulting in the increase of the freezing/melting points.\cite{Kaneko:cpl:2010}
Therefore, a larger $H$ value is selected in this study to reflect practical experimental conditions.
Subsequently, the constraint of the $z$-position of the upper mica is released, surface pressure is applied, and the time evolution of the $z$-position of the upper mica is measured until a plateau is reached.
Lastly, the OMCTS molecules outside the mica walls were removed from the system, and the equilibrium simulations were repeated.
Confined systems with a targeted number of layers ($n$ = 2, 3, and 5) were obtained using this procedure, and the equilibrium properties of the system are explained in Section~\ref{subsec:equilibrium}.

After performing the equilibrium simulations, shear flow was imposed by applying a constant sliding velocity, $V_{\rm x}$, on the upper mica.
The fluid temperature was controlled at 300\,K by applying the velocity scaling method to the thermal velocity, which was subtracted from the streaming velocity.
The applied $V_{\rm x}$ ranged from $10^{-3}\,{\rm m/s}$ to $10^{2}\,{\rm m/s}$, leading to shear rates, $\dot{\gamma}$, in the range from $1.32\times 10^{5}\,{\rm s^{-1}} \lesssim \dot{\gamma} \lesssim 3.57\times 10^{10}\,{\rm s^{-1}}$ for the film thickness simulated.
The given low sliding velocities ($V_{\rm x} < 10^{-2}\,{\rm m/s}$) are of particular interest because this range enables an overlap with those used in the tribometer experiments.
Conversely, there are significant fluctuations in friction force and velocity gradient at such low $V_{\rm x}$ owing to the thermal noise. Furthermore, the calculation time required for the simulation to reach a steady state is also very long.
Therefore, a moving average was applied to smooth out the data to determine the steady state and the shear viscosity, $\eta$, which is given by:
\begin{equation}
	\eta = \tau\frac{H}{V_{\rm x}}
\end{equation}
where $\tau$ denotes the shear stress, which is measured by the friction force on the upper mica per unit surface area.
We used the graphics processing unit (GPU) versions\cite{Brown:cpc:2011} of the large-scale atomic/molecular massively parallel simulator (LAMMPS)\cite{Plimpton:jcp:1995} MD package are used in this study.

\section{Results}
\label{sec:results}
\subsection{Equilibrium behavior}
\label{subsec:equilibrium}
\begin{figure}[t]
	\centering
	\includegraphics[width=8cm]{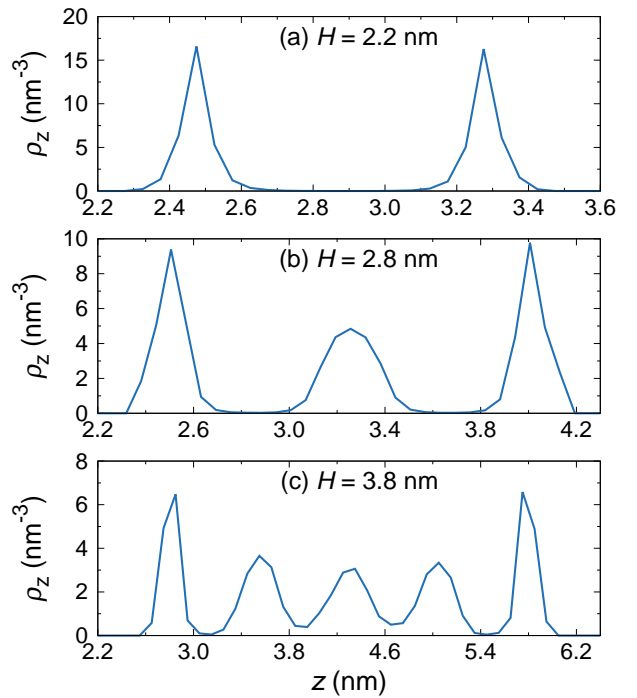}
	\caption{Density profiles of confined OMCTS in the direction normal to the mica surface for different surface separations.}
	\label{fig:DensProf}
\end{figure}
\begin{table}[b]
\caption{Equilibrium properties of confined OMCTS}
\begin{center}
  \begin{tabular*}{0.48\textwidth}{@{\extracolsep{\fill}}cccc}
    \hline\hline
    $H$\,(nm) & $P_{\rm zz}$\,(MPa)\cite{Mizukami:private:2022} & $\rho$\,(nm$^{-3}$) & $D \times 10^{-9}$\,(m$^2$/s)\\
    \hline\hline
    2.2  & 2.0 & 1.41 & 0.045\\
    2.8  & 1.0 & 1.48 & 0.177\\
    3.8  & 0.1 & 1.63 & 0.107\\
    \hline
  \end{tabular*}
  \end{center}
  \label{tab:EquilProp}
\end{table}
We investigated the equilibrium properties of the confined systems with layered structures ($n$ = 2, 3, and 5).
Figure~\ref{fig:DensProf} presents the density profiles of the centers-of-mass of the OMCTS along the direction normal to the mica surface for different surface separations.
We can clearly observe the layered structures can be clearly observed in the confined system for all values of $H$.
Zero density is observed between the layers at $H=2.2$ ($n=2$) and 2.8 ($n=3$).
However, a non-zero density is observed at $H=3.8$ ($n=5$), indicating a loose layered structure.
Previously conducted simulation studies\cite{Matsubara:jcp:2011,Matsubara:prl:2012,Vadhana:jpcb:2016} have reported similar behavior.
Table~\ref{tab:EquilProp} lists the resulting equilibrium properties for each $H$ value, including the surface pressure $P_{\rm zz}$, number densities $\rho_{\rm z}$, and diffusion coefficient $D$.
In a previously conducted molecular simulation study of the OMCTS-mica system\cite{Matsubara:prl:2012}, the reason for selecting a specific $H$ within a single layer region was ambiguous.
We used the value of $P_{\rm zz}$ obtained from the surface force measurements\cite{Mizukami:private:2022} to reproduce the realistic state of the confined system.
These settings can be used to capture phenomena in practical systems more accurately, and the obtained results can be connected with the behavior observed in experiments.
The diffusion coefficient, $D$, was calculated from the slopes of the time-averaged mean-square displacement curves.
A previous study\cite{Matsubara:prl:2012} reported a diffusion slowdown induced by confinement, where $D$ was approximately two orders of magnitude lesser than the bulk value\cite{Matsubara:jctc:2010}.
A considerable decrease was observed for $H=2.2\,{\rm nm}$, which was attributed to the contact layers.
For a confined system, the existence of contact layers causes a significant decrease in the diffusivity.\cite{Matsubara:prl:2012}
It is observed that $D$ increased despite the decrease of the surface separation from $H=3.8\,{\rm nm}$ to $H=2.8\,{\rm nm}$.
Previous simulations \cite{Matsubara:prl:2012} have also reported that the diffusion curve exhibits a non-monotonous decrease when considered as a function of surface separation.
We found that this result is primarily attributed to the number density of the confined systems (see Table~\ref{tab:EquilProp}).

\begin{figure}[tbp]
	\centering
	\includegraphics[width=8cm]{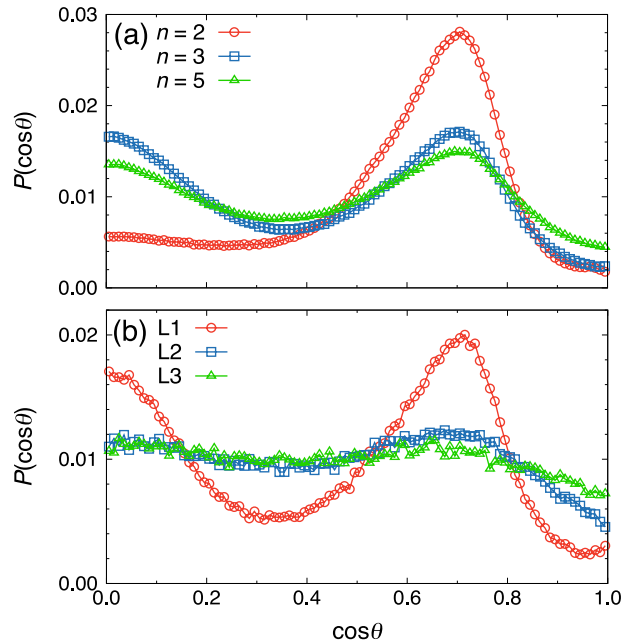}
	\caption{(a) Orientational distributions, $P({\rm cos}\,\theta)$, of confined OMCTS for various surface separations, $H$.
	$\theta$ denotes the angle between the Si and Si vector and the normal to the mica surface. (b) Layer-wise analysis of 
	$P({\rm cos}\,\theta)$ for $H=3.8\,{\rm nm}$ ($n=5$). L1 is the contact layer, L2 is the second contact layer, and 
	L3 is the middle layer.}
	\label{fig:OrientEquil}
\end{figure}
The orientational distributions, $P({\rm cos}\,\theta)$, of the confined OMCTS, where $\theta$ is the angle between the Si and Si vector and normal to the mica surface, are computed to obtain the detailed structural information.
A perpendicular orientation with the surface normal corresponds to ${\rm cos}\,\theta=0$ and a parallel orientation with the surface normal corresponds to ${\rm cos}\,\theta=1.0$.
Figure~\ref{fig:OrientEquil}(a) presents $P({\rm cos}\,\theta)$, of confined OMCTS for various surface separations.
For $H=2.2\,{\rm nm}$, a distinct peak is observed at ${\rm cos}\,\theta \approx 0.7$.
This specific orientation was also observed in the previous simulation, but the distribution shifted to ${\rm cos}\,\theta = 0$ when the surface separation was decreased further.\cite{Vadhana:jpcb:2016}
For larger separations with contact layers and middle layers ($H=2.8\,{\rm nm}$ $(n=3)$ and $H=3.8\,{\rm nm}$ $(n=5)$), 
we observed a different behavior, where two distinct peaks appeared at ${\rm cos}\,\theta = 0$ and ${\rm cos}\,\theta \approx 0.7$.
The peak at ${\rm cos}\,\theta \approx 0.7$ gradually weakens when compared to the case of $H=2.2\,{\rm nm}$.
A broader distribution was observed with the increase of $H$, indicating that the OMCTS molecules did not have any preferred orientation corresponding to the surface normal.
A layer-by-layer analysis of $P({\rm cos}\,\theta)$ is performed for $H=3.8\,{\rm nm}$ $(n=5)$ to obtain a more detailed understanding of the difference between $n=2$ and others, as shown in Fig.~\ref{fig:OrientEquil}(b).
It can be observed from this plot, that the first contact layers (L1) had sharper peaks when compared to the second layer (L2) and middle layer (L3).
The peaks clearly became weak for L2 and L3, and a more uniform distribution is observed in the central region (from L2 to L3).
The formation mechanism of the layered structures in contact layers that have two distributions is attributed to the molecular pair of contact layers and inner layers induced by the confinement effect, as explained by a previous study conducted by Matsubara $et$ $al.$\cite{Matsubara:jcp:2011}.
These characteristic structures induced by nanoconfinement significantly affect both the equilibrium properties and the flow properties, which will be discussed in the following section.

\subsection{Flow behavior}
\begin{figure}[bp]
	\centering
	\includegraphics[width=8cm]{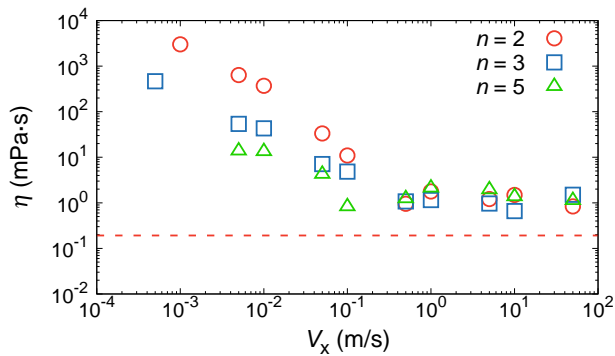}
	\caption{Shear viscosity, $\eta$, as a function of sliding velocity, $V_{\rm x}$, on the upper mica. The dashed red line represents the
	zero-shear viscosity for $H=2.2\,{\rm nm}$, which is determined from the Green-Kubo relation.}
	\label{fig:ShearVisc}
\end{figure}
This section presents a detailed description of the structure formation and viscosity behavior to elucidate the underlying physical phenomena of flow behavior for low sliding velocities, reflecting practical operating conditions.
Figure~\ref{fig:ShearVisc} illustrates the shear viscosity, $\eta$, of OMCTS confined between two mica surfaces as a function of the sliding velocity, $V_{\rm x}$, on the upper mica.
For a relatively low sliding velocity, $V_{\rm x}\lesssim 10^{-1}\,{\rm m/s}$, the confined liquids exhibited a distinct shear-thinning behavior for all the systems.
When a higher sliding velocity, $V_{\rm x}\gtrsim 10^{-1}\,{\rm m/s}$, was applied, a Newtonian-like behavior was observed with $\eta \approx$ constant which corresponds to a slightly smaller zero-shear viscosity of the bulk liquid, $\eta_{\rm bulk}=2.13\,{\rm mPa \cdot s}$.
Additionally, it was observed that the shear viscosities at $V_{\rm x}\lesssim 10^{-1}\,{\rm m/s}$ were largely dependent on the number of layered structures, $n$.
$\eta$ increases with the increase in $H$, particularly for $H=2.2$ ($n=2$).

\begin{figure}[t]
	\centering
	\includegraphics[width=8cm]{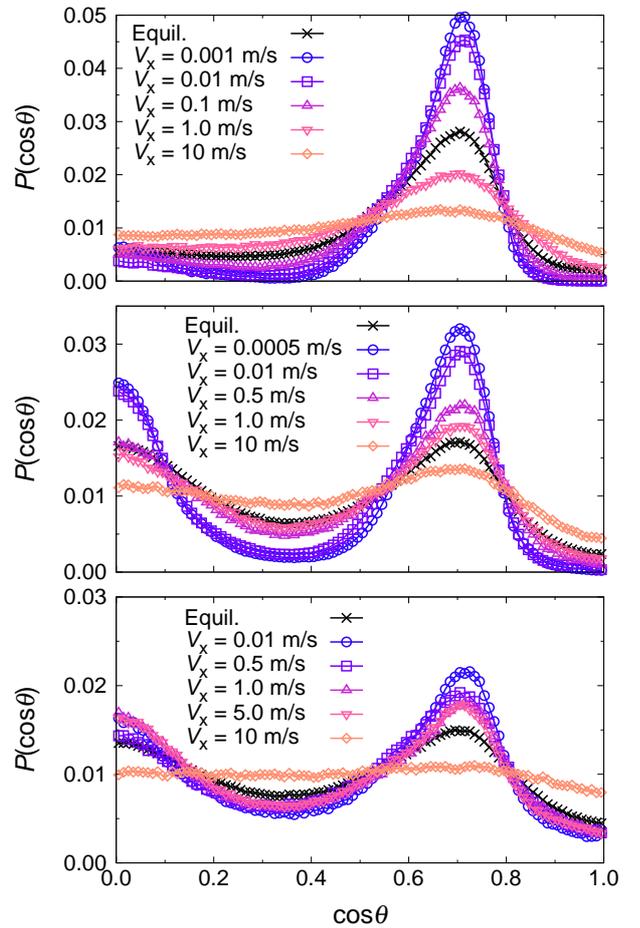}
	\caption{Orientational distributions, $P({\rm cos}\,\theta)$, of confined OMCTS at (a) $H=2.2\,{\rm nm}$, (b) 
	$H=2.8\,{\rm nm}$, and (c) $H=3.8\,{\rm nm}$ for various $V_{\rm x}$. For comparison purposes, the corresponding 
	data in equilibrium are represented as `Equil.'.}
	\label{fig:OrientShear}
\end{figure}
The orientational distributions of the confined OMCTS under shear, $P({\rm cos}\,\theta)$, are analyzed to better understand the mechanism behind this viscosity change, as shown in Fig.~\ref{fig:OrientShear}.
The shear-induced rearrangements, such as enhancement of the orientation degree and breakup of the oriented structures, are observed for all $H$ values.
It can be observed that a relatively weak shear ($V_{\rm x}\lesssim 10^{-1}\,{\rm m/s}$) observed at $H=2.2\,{\rm nm}$ ($n=2$) leads to an increase in the distribution at ${\rm cos}\,\theta \approx 0.7$ (see Fig.~\ref{fig:OrientShear}(a)).
This is interesting because of a possible scenario for the increase in shear viscosity at low sliding velocities, which reflects the actual operating conditions.
Additionally, the zero-shear viscosity data is also plotted for $H=2.2\,{\rm nm}$ using the Green-Kubo relation in Fig.~\ref{fig:OrientShear}, as follows:
\begin{equation}
	\eta_0 = \frac{V}{k_{\rm B}T}\int_{0}^{\infty}\la P_{\alpha\beta}(0)P_{\alpha\beta}(t)\ra dt
	\label{eq:GK}
\end{equation}
where $P_{\alpha\beta}$ denotes the off-diagonal components of the pressure tensor.
While applying the Green-Kubo relation to confined systems is a controversial debate\cite{Kohler:pccp:2017,Mart:ent:2017,Zaragoza:pccp:2019} because of the inability to adapt to heterogeneous systems, we computed the zero-shear viscosity in parallel (defined by axial pressure components $P_{\rm xy}$) and perpendicular (defined by radial components $P_{\rm xz}$ and $P_{\rm yz}$) directions and then averaged them.
The viscosity computed using the Green-Kubo formula in this study was almost an order of magnitude smaller than the bulk viscosity, which is similar to that reported in a previous study\cite{Zaragoza:pccp:2019}.
The calculated results cannot be guaranteed in a confined system since both the liquid-liquid interactions and the solid-liquid boundary condition affect the thermal fluctuations of the shear stress in a confined liquid.
However, the layered structures under low $V_{\rm x}$ have a strong degree of orientation at ${\rm cos}\,\theta \approx 0.7$ when compared to that in an equilibrium state, as shown in Fig.~\ref{fig:OrientShear}(a); the shear-induced rearrangement can cause an enhanced shear viscosity.
It will be interesting to reveal the correlation between the shear viscosity and structure formation, such as the degree of order under extremely low sliding velocities ($V_{\rm x}<10^{-3}\,{\rm m/s}$), in future work.
Additionally, the development of theoretical prediction methods, as well as the issues of both large thermal noise and high computational cost to reach a steady state, will be the objective of future studies.
We continue our discussion with the confined systems for $n>2$.
An increase in the distribution at ${\rm cos}\,\theta \approx 0.7$ was observed at $H=2.8\,{\rm nm}$ ($n=3$), even for a relatively high sliding velocity, $V_{\rm x}=10^{0}\,{\rm m/s}$.
In addition to the peak at ${\rm cos}\,\theta \approx 0.7$, an increase in the distribution is observed at ${\rm cos}\,\theta = 0$; however, the peak intensity at ${\rm cos}\,\theta = 0$ was weaker than that at ${\rm cos}\,\theta \approx 0.7$ (see Fig.~\ref{fig:OrientShear}(b)).
The same behavior was observed at $H=3.8\,{\rm nm}$ ($n=5$) as in the case of $n=3$, but $\eta$ was smaller than that in the case of $n=3$ for the shear-thinning regime ($V_{\rm x}<5.0\times 10^{-1}\,{\rm m/s}$).
The three-layered system had a higher $\eta$ value for the shear-thinning regime ($V_{\rm x}<5.0\times 10^{-1}\,{\rm m/s}$) than that in the case of $n=3$ although the self-diffusion coefficient, $D$, of the three-layered system was higher than that of the five-layered system (see Table~\ref{tab:EquilProp}).
This indicates that the formation of oriented structures in a confined liquid is one of the factors which significantly affects the shear viscosity.

Subsequently, the origin of the shear-thinning of the confined liquids is discussed.
A clear correlation can be obtained between the gradual breakup of the oriented structures and shear-thinning behavior by comparing Fig.~\ref{fig:ShearVisc} and Fig.~\ref{fig:OrientShear}.
This can be more clearly observed from the result of $H=2.2\,{\rm nm}$.
The intensity of the peak at ${\rm cos}\,\theta \approx 0.7$ gradually decreased along with shear-thinning as the sliding velocity increased to $V_{\rm x}=10^{-1}\,{\rm m/s}$.
The confined liquid exhibits a Newtonian-like behavior even though the peak at ${\rm cos}\,\theta \approx 0.7$ remains at $V_{\rm x}=10^{0}\,{\rm m/s}$.
We anticipate the existence of a threshold peak intensity below which there is no significant effect on the viscosity behavior of confined liquids.
For larger separations ($n>2$), the slope of $\eta$ becomes less steep when compared to the case of $n=2$, as shown in Fig.~\ref{fig:OrientShear}.
The effect of the specific orientation distribution of the contact layers on the shear viscosity reduces with the increase in $H$ because of the existence of middle layers which have random orientations.
From the peak intensity at ${\rm cos}\,\theta \approx 0.7$ for all the analyzed $H$ (Fig.~\ref{fig:OrientShear}(a-c)) values, the threshold for the molecular orientation which causes the shear-thinning of the confined liquids was observed to be approximately 2\,\%.
When $V_{\rm x}$ was further increased ($V_{\rm x}\gtrsim 10^{0}\,{\rm m/s}$), we observed a broader orientational distribution for all the $H$ values, indicating that the OMCTS molecules do not have any preferred orientation.
These confined liquids continued to exhibit a Newtonian response to shear while breaking up the oriented structures.
The correlation between the shear viscosity and structure formation corresponds to all the analyzed $H$ values, and the obtained results indicate that the shear viscosity of simple nonpolar lubricants under nanoslit confinement is primarily determined by the degree of molecular orientation.

\section{Conclusions}
\label{sec:conclusions}
We studied the correlation between the shear viscosity, $\eta$, and the structure formation of OMCTS confined between two mica walls using the MD method.
Three different surface separations, $H$, reproducing two- ($n=2$), three- ($n=3$), and five-layered ($n=5$) structures, were prepared by performing liquid-vapor molecular dynamics simulations.
A confinement-induced diffusion slowdown is observed at rest, particularly for $n=2$.
For larger separations ($n>2$), the self-diffusion coefficient, $D$, of $n=5$ was higher than that of $n=3$, and we found that this was because of the difference in the number density of the confined systems.
In addition to the transport properties, the orientational distributions of the confined OMCTS molecules were also analyzed.
We observed distinct distribution with one specific peak for $n=2$ and two distributions, including a parallel orientation with the surface normal for $n>2$.
These distributions for $n>2$ became more uniform when moving towards the central region in the confined system.

We then applied sliding velocities to the upper mica wall and generated shear flow covering the actual operating conditions in the range of $10^{-3}\,{\rm m/s} \le V_{\rm x} \le 10^{2}\,{\rm m/s}$.
The confined liquids demonstrated a distinct shear-thinning behavior for all the analyzed $H$ values for a relatively low sliding velocity, $V_{\rm x}\lesssim 10^{-1}\,{\rm m/s}$. Furthermore, it is observed that $\eta$ at $V_{\rm x}\lesssim 10^{-1}\,{\rm m/s}$ depends on the number of layered structures, contrary to the results of the diffusion coefficient in equilibrium.
Newtonian behavior was observed when a higher sliding velocity, $V_{\rm x}\gtrsim 10^{-1}\,{\rm m/s}$, was applied.
A strong correlation was identified between the degree of molecular orientation and shear viscosity of confined liquids.
By applying a low $V_{\rm x}$, we observed the shear-induced enhancement of orientation degree compared to the equilibrium state. The magnitude of $\eta$ in the shear-thinning regime was found to be dependent on the orientation intensity.
Furthermore, it was observed that shear-thinning originated from the vanishing of specific orientational distributions, corresponding to the gradual breakup of the oriented structures.
The findings of this study provide new insights into the correlation between the structural and lubrication properties in the mixed/boundary lubrication regime and support breaking away from the trial-and-error empirical approaches for developing and improving lubricants.

\section*{Acknowledgements}
This work was supported in part by the Ministry of Education, Culture, Sports, Science and Technology (MEXT) as Research and Development of Next-Generation Fields. The computation was partially carried out using the computer resources offered under the category of general projects by the Research Institute for Information Technology, Kyushu University. The authors are grateful to Prof. K. Kurihara (Tohoku University) and Prof. M. Mizukami (Tohoku University) for helpful discussions and useful comments.


\bibliography{cite} 

\clearpage

\onecolumngrid
{
    \center \bf \large 
    Supplementary Materials for \\``Correlation between ordering and shear thinning in confined liquids''\vspace*{1cm}\\ 
    \vspace*{0.0cm}
}

\renewcommand{\figurename}{FIG.~S\hspace{-0.25em}}
\setcounter{figure}{0}

\begin{figure}[h]
	\centering
	\includegraphics[width=14cm]{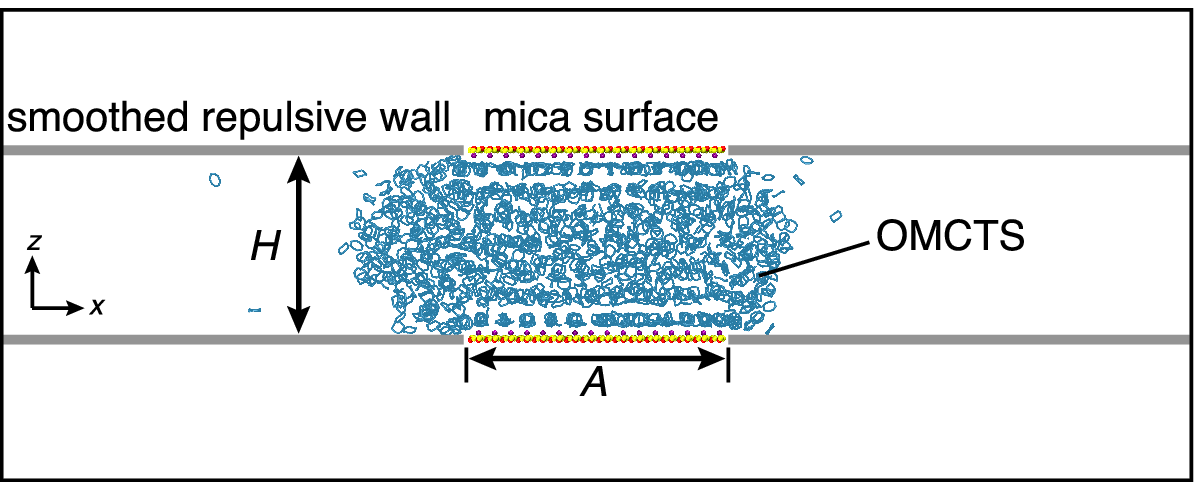}
	\caption{Mica-octamethylcyclotetrasiloxane (OMCTS) system for the molecular dynamics simulation with liquid-vapor ensemble.}
	\label{SMfig:LVMD}
\end{figure}

\begin{figure}[b]
	\centering
	\includegraphics[width=14cm]{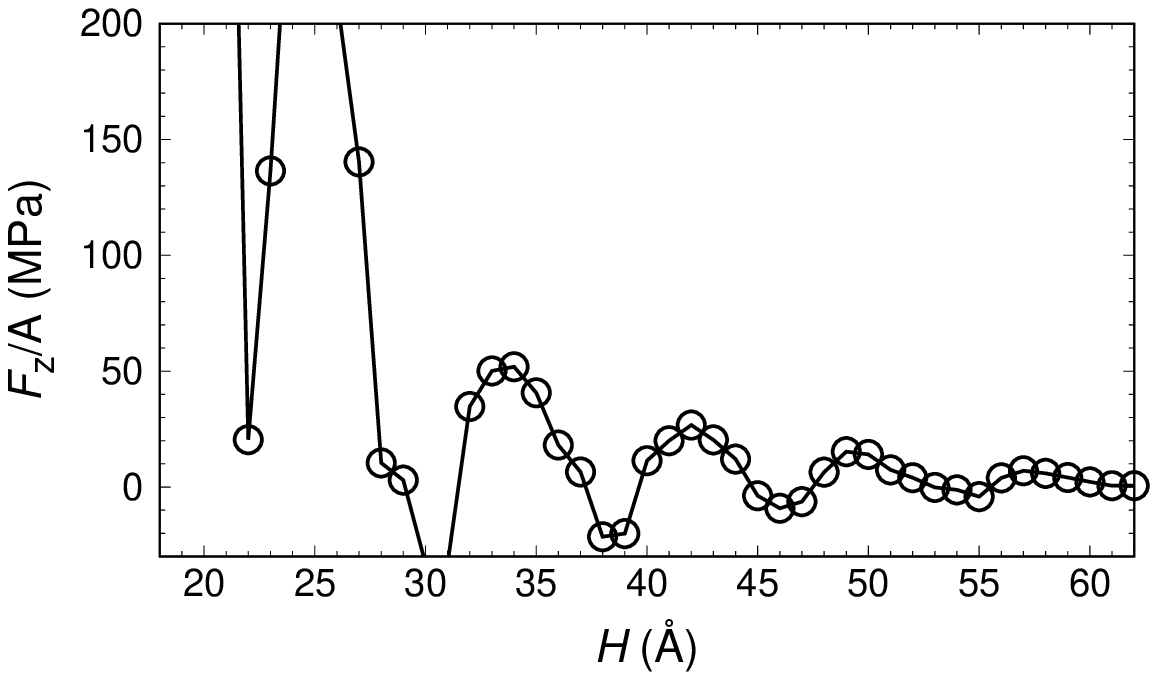}
	\caption{Force curve calculated from the molecular dynamics simulation with liquid-vapor ensemble.}
	\label{SMfig:Fcurve}
\end{figure}

\end{document}